\documentclass[useAMS,usenatbib]{mn2e}
\usepackage{amssymb}
\usepackage{graphicx}
\begin{document}

\title{Spectral features in isolated neutron stars induced by inhomogeneous surface temperatures}


\author[D.~Vigan\`o et al.]{Daniele Vigan\`o$^1$, Rosalba~Perna$^2$,
  Nanda~Rea$^{1,3}$, Jos\'e~A.~Pons$^4$\\ $^1$ Institute of Space
  Sciences (CSIC--IEEC), Campus UAB, Faculty of Science, Torre
  C5-parell, 08193 Bellaterra, Spain\\ $^2$ Department of Physics and
  Astronomy, Stony Brook University, Stony Brook, NY, 11794,
  USA\\ $^3$ Anton Pannekoek Institute, University of Amsterdam, Postbus 94249,
1090GE Amsterdam, The Netherlands\\ $^4$
  Departament de F\'isica Aplicada, Universitat d'Alacant, Ap. Correus
  99, 03080 Alacant, Spain}

\date{}
\maketitle

\label{firstpage}

\begin{abstract}
The thermal X-ray spectra of several isolated neutron stars display
deviations from a pure blackbody. The accurate physical interpretation
of these spectral features bears profound implications for our
understanding of the atmospheric composition, magnetic field strength
and topology, and equation of state of dense matter. With specific
details varying from source to source, common explanations for the
features have ranged from atomic transitions in the magnetized
atmospheres or condensed surface, to cyclotron lines generated in a
hot ionized layer near the surface. Here we quantitatively evaluate
the X-ray spectral distortions induced by inhomogeneous temperature
distributions of the neutron star surface. To this aim, we explore
several surface temperature distributions, we simulate their
corresponding general relativistic X-ray spectra (assuming an
isotropic, blackbody emission), and fit the latter with a single
blackbody model. We find that, in some cases, the presence of a
spurious 'spectral line' is required at a high significance level in
order to obtain statistically acceptable fits, with central energy and
equivalent width similar to the values typically observed. We also
perform a fit to a specific object, RX~J0806.4-4123, finding several
surface temperature distributions able to model the observed
spectrum. The explored effect is unlikely to work in all sources with
detected lines, but in some cases it can indeed  be responsible for the appearance of such lines. Our results enforce the idea that surface
temperature anisotropy can be an important factor that should be
considered and explored also in combination with more sophisticated
emission models like atmospheres.
\end{abstract}

\section{Introduction}

Isolated neutron stars (NSs) display an observational variety of timing
and spectral properties which has led, historically, to a
classification in different subclasses. Rotation-powered pulsars
(RPPs, \citealt{becker09}) constitute the majority of known pulsars;
they convert part of their rotational energy into non-thermal
radiation across the whole electromagnetic spectrum. More than two
thousand sources are visible in radio, and more than one hundred in
X-rays and/or $\gamma$-rays. Other NSs are detected only or mostly in
X-rays: $\sim 20$ magnetars \citep{mereghetti08}, the seven nearby
thermally emitting NSs (simply named X-ray Isolated NSs, XINSs,
\citealt{turolla09}) and the handful of heterogeneous central compact
objects (CCOs, \citealt{gotthelf13}).

In about 40 of the pulsars visible in X-rays, the spectrum shows
evidence for thermal emission (see \citealt{vigano13} and our
website\footnote{{\tt www.neutronstarcooling.info}} for a
catalog). Therefore, the X-ray spectra carry precious information about
the surface properties and the physics of the crust. However, the
ability to draw firm implications from the analysis of the properties
of the thermal emission is often hampered by the faintness of some
objects, the co-existence of different physical mechanisms
contributing to the detected radiation (rotation-powered non-thermal
radiation in RPPs, and magnetospheric Compton scattering in
magnetars), and the presence of interstellar absorption.

The temperatures inferred from blackbody (BB) fits to X-ray spectra
range from $kT_{bb}\sim 45$-$100$ eV (in XINSs and some standard
radio-pulsars), up to 300-500 eV in some magnetars (during their
quiescent state). The inferred radii are also systematically smaller
than the typical 10-12 km expected from a NS. In some cases, e.g. the
three musketeers \citep{deluca05}, the CCO in Puppis A
\citep{gotthelf10,deluca12}, and a few magnetars
\citep{halpern05,bernardini09,albano10,bernardini11}, phase-averaged
and phase-resolved spectra can be modelled by considering spectral models with two or three
temperatures, physically interpreted as coming from different 
regions of the star surface.

\begin{table*}
\begin{center}

\caption{Broad spectral features in the $X$-ray spectra of isolated
  NSs. We report the best-fit parameters for the best BB+line model
  (where line model is usually {\tt gabs, gauss} or {\tt cyclabs}) and the pulsed fraction PF, as
  found in literature (when more than one reference is found, we
  consider the observation with the largest number of photons). We
  also report the estimated surface dipolar magnetic field (at the
  pole), as inferred from the timing properties: $B_{dip}=6.4\times
  10^{19}(P\mbox{[s]}\dot{P})^{1/2}$~G.}
\label{tab:data_lines}
\begin{tabular}[t!]{l c c c c c c c c}
\hline 
\hline 
Source 				& Class & $B_{dip}$ & $N_H$ 	& $kT_{bb}$ & $E_0$ & $|E_w|$	& PF & Refs.	\\
	& 			& [$10^{12}$G] & [$10^{20}{\rm cm}^{-2}$] 	& [eV] & [eV] 		& [eV] &$\%$	& \\
\hline 
RX J0720.4-3125		 	& XINS 	& 49		& 1.0		& 84-94		& 311$^\star$	& 0-70		& 11 & [1] 	 \\
RX J0806.4-4123 			& XINS	& 51		& 0.9		& 95			& 486$^\star$	& 30			& 6 & [2] 	  \\	
RX J1308.6+2127 			& XINS	& 68		& 3.7		& 93			& 390$^\star$	& 150		& 18 & [3] 	 \\
RX J1605.3+3249		 	& XINS	& 148$^\dagger$& 0	& 99			& 400$^\star$	& 70			& 5$^\dagger$ & [4] 		\\
RX J2143.0+0654			& XINS	& 40		& 2.3		& 104		& 750	 	& 50			& 4 & [5] 		\\
2XMM J1046-5943$^\ddagger$	& ?		& ?		& 26		& 135		& 1350$^\star$	& 90			& $<$4& [6]		\\
1E 1207.4-5209			& CCO	& 0.2 	& 13		& 155,290 	& 740,1390	& 60,100		& 4-14$^{**}$ & [7] 	\\
PSR J1740+1000		& RPP	& 37		& 9.7 	& 94			& 550-650  	& 50-230		& 30 & [8]  \\
PSR J1819-1458	 		& RPP	& 100 	& 124	& 112		& 1120$^\star$ 	& 400		& 34 & [9] 	 \\
XTE J1810-197			& MAG	& 410	& 73		& 300		& 1150		& 35			& 17-47$^{**}$ 	& [10]		\\
\hline
\hline
\end{tabular}
\end{center}

\begin{minipage}{\textwidth}
$^\star$ The best-fit model consists of additional lines and/or a second BB. Values for the BB+line fit are shown.\\
$^\dagger$ To be confirmed: $P$ and $\dot{P}$ values found at $4\sigma$ and $2\sigma$ levels \citep{pires14}.\\
$^\ddagger$ No pulsations detected so far.\\
$^{**}$ Depending on the energy range.\\
References: [1] \cite{devries04}; \cite{haberl04b}; \cite{haberl06}; \cite{hohle12}; [2] \cite{haberl04}; \cite{kaplan09b}; [3] \cite{haberl03}; \cite{schwope07}; \cite{hambaryan11}; [4] \cite{vankerkwijk04}; \cite{pires14}; [5] \cite{zane05}; \cite{kaplan09}; [6] \cite{pires12}; [7] \cite{sanwal02}; \cite{bignami03}; \cite{deluca04}; \cite{mori05}; [8] \cite{kargaltsev12}; [9] \cite{mclaughlin07}; \cite{miller13}; [10] \cite{bernardini09}.
\end{minipage}
\end{table*} 


Other objects, and in particular the XINSs, display a spectrum which
is often broadly consistent with a single-temperature model. However,
when the source is observed at a sufficiently high signal-to-noise,
deviations are seen, often in the form of a feature in the
spectrum. 

Deviations from a pure blackbody spectrum can be produced by several
physical effects. Firstly, the presence of a thin atmospheric layer
\citep{romani87,vanriper88,miller92,pavlov95,zavlin98,ho01,ozel01,ho03a,ho03b,ho03c,mori06,
  vanadelsberg06,suleimanov09,suleimanov10,suleimanov12} distorts and
broadens the spectrum.  The (poorly known) chemical composition of the
atmosphere and the degree of ionization of the species, the strong
magnetic field, the vacuum polarization effect, all introduce various
distortions and features in the resulting spectra.  Also, as an
alternative to atmospheres, condensed surface models have been
proposed \citep{vanadelsberg05,perezazorin06b}, and could be
particularly suitable for strongly magnetized objects, in which the
outer layers consist of atomic chains \citep{chen74}. Hybrid models,
with a condensed surface covered by a thin atmosphere, have been
proposed to explain the optical to X-ray spectra of the XINS RX
J1856.5-3754 \citep{ho07,ho07b}.

Spectral features in the spectra can be produced by proton and electron
cyclotron emission/absorption at harmonics of the cyclotron frequency,
and by electronic transitions in neutral and partially ionized atoms
\citep{sanwal02,vankerkwijk04,suleimanov12}. However, at high magnetic
fields, $B\gtrsim 10^{14}$~G, vacuum polarization tends to reduce the
contrast of those spectral features \citep{ho03b}.

Spectral distortions from a pure BB are expected also as a result of
an inhomogeneous temperature distribution on the NS surface. Such
temperature inhomogeneities can be inferred from measurements of small
emitting BB areas in a number of isolated NSs
(e.g. \citealt{halpern10}), and are theoretically predicted, e.g., by
anisotropic thermal conductivity induced by the presence of strong magnetic
fields (e.g. \citealt{page07,pons09}), and by magnetospheric particle
bombardment \citep{cheng80}.

In this work, we perform a general study of the resulting spectral features arising from different surface temperature distributions. In the
spirit of focusing on this effect only, and of separating its
contribution from the other effects discussed above, we consider a
local isotropic BB emission; however, the same concept could be
applied to more sophisticated emission models. We note that our study is of relevance to the cases in which
the spectral feature is detected in the thermal part of the spectrum, and
when it is relatively broad. 

For the purpose of our investigation, we consider several
representative axisymmetric surface temperature distributions, we
numerically compute their general relativistic spectra (both
phase-averaged and phase-resolved), and we use them to simulate
'observational' data.  We then fit these synthetic spectra with a BB
model and a BB plus an absorption line model, and we explore the
conditions under which the presence of a 'spurious line' in the spectra is
required at high significance in the simulated spectra. We then consider the actual {\em XMM-Newton} data for a
particular source, RX~J0806.4-4123, for which an absorption line has
been claimed in the spectrum. We show that it is possible to find
several temperature profiles which provide a good overall fit to the
observed spectral distribution.

The paper is organized as follows: in \S~\ref{sec:obs}, we review the
literature on absorption features in NS X-ray spectra. In
\S~\ref{sec:synthetic}, we present the considered axisymmetric surface
temperature distributions, and calculate the corresponding
general-relativistic spectra. In \S~\ref{sec:results}, we compute
synthetic spectra from the theoretical ones, and discuss the fit
results to these synthetic spectra when modeled with a BB and a BB
plus an absorption line. We further present the results of the fit to
RX~J0806.4-4123. Finally, we discuss and summarize our results in
\S~\ref{sec:discussion}.

\begin{figure*}
\centering
\includegraphics[width=.33\textwidth]{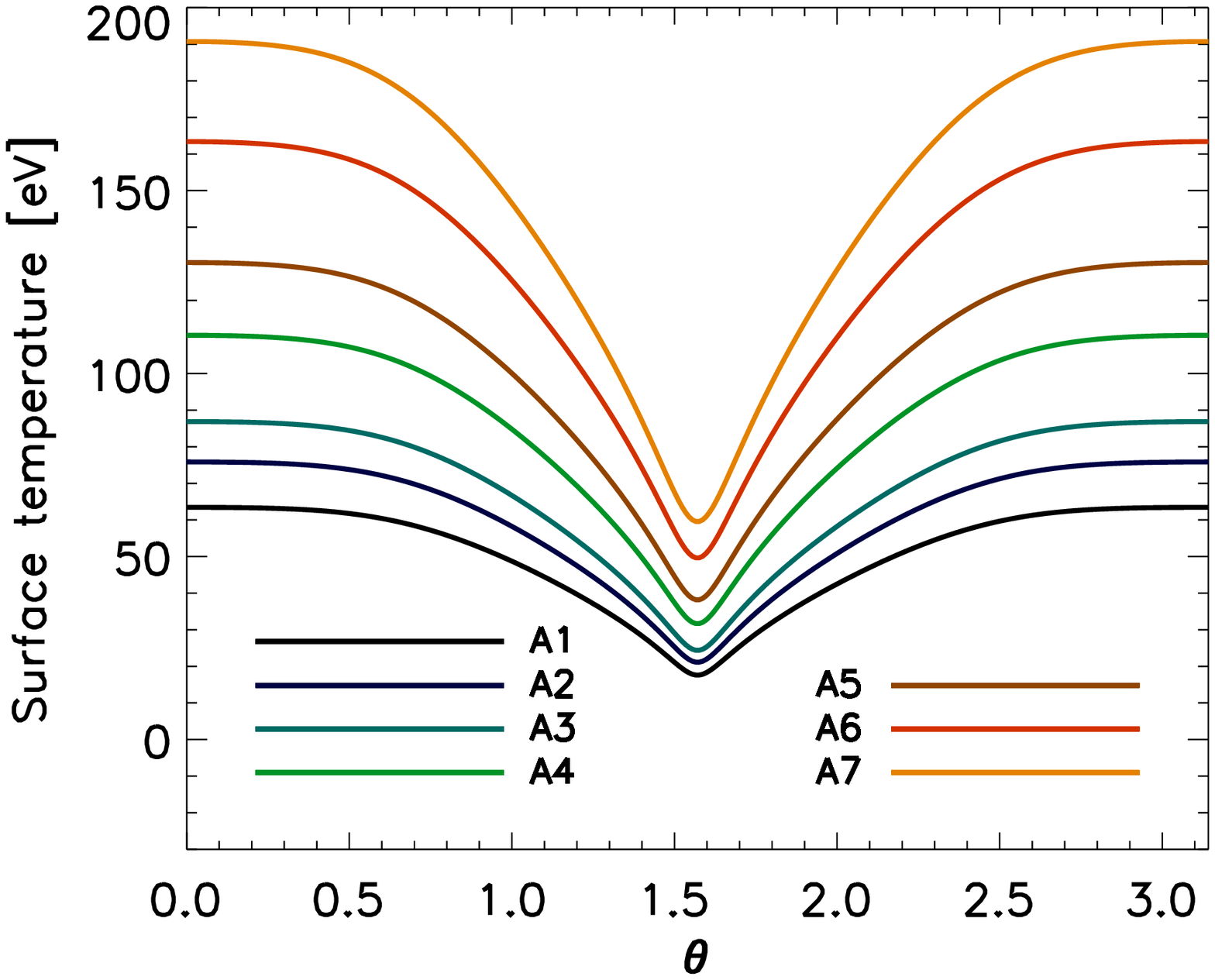}
\includegraphics[width=.33\textwidth]{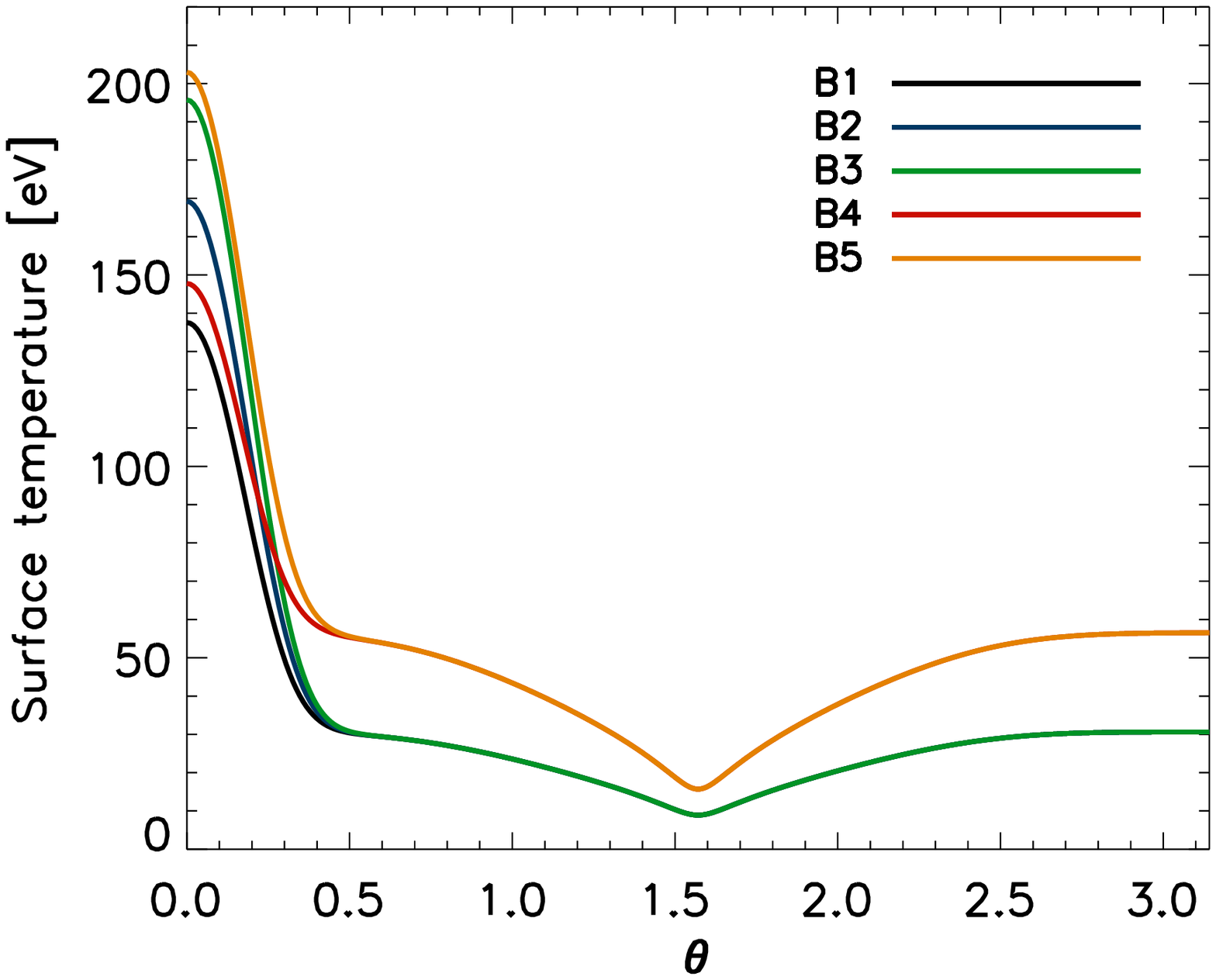}
\includegraphics[width=.33\textwidth]{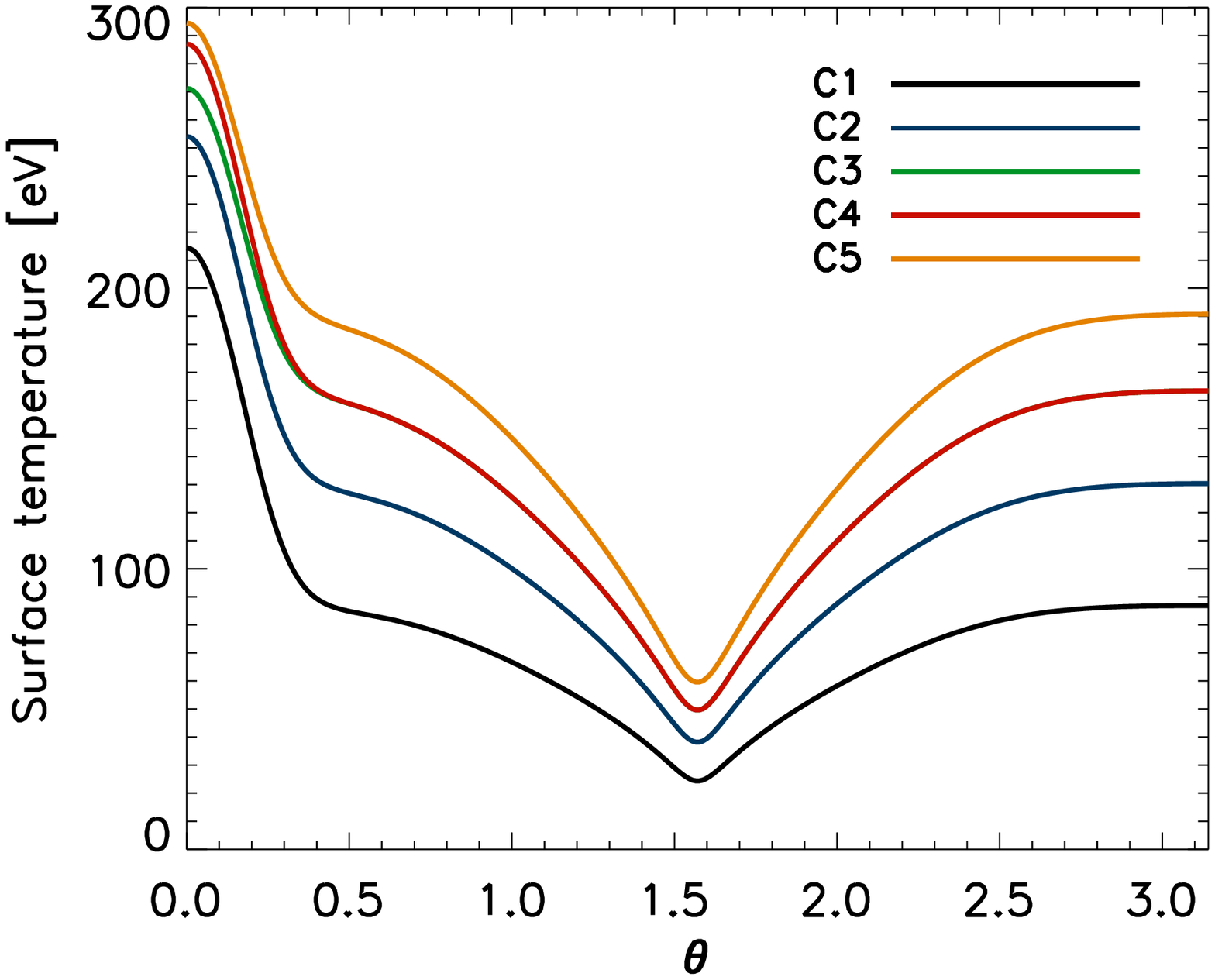}
\caption{Surface (unredshifted) temperatures of the families of models A, B, C (left, middle, right panels, respectively).}
 \label{fig:temperatures}
\end{figure*}

\section{Observed spectral features}\label{sec:obs}

In Table~\ref{tab:data_lines} we list the reported broad spectral
features in isolated neutron stars (XINSs, CCOs, RPPs, magnetars).
Most of the reported cases resulted from the analysis of {\em
  XMM-Newton}/EPIC-pn data, thanks to its large effective
area.\footnote{\cite{hohle12b}, by using the spectrometer {\em
    XMM-Newton}/RGS, found several very narrow features in nearby NSs,
  whose origin could also be due to absorption by the interstellar
  medium. We do not consider these features here, since they are too
  narrow and cannot be produced by an inhomogeneous temperature
  distribution.}

The XINSs are seven nearby, middle-aged ($\sim 10^5-10^6$ yr), and slowly
rotating neutron stars. Their surface X-ray emission is not
reprocessed by a dense magnetospheric plasma (as in magnetars) or
contaminated by rotationally-powered non-thermal photons (as in RPPs);
also, since they are nearby objects, they suffer little
interstellar absorption ($N_H \sim 10^{20}$ cm$^{-2}$).  As a
consequence, the XINSs are the ideal candidates to study the NS
surface emission properties. Their deepest observations have collected
$\sim 10^4-10^5$ photons, most of which at low energy (below 1 keV),
and with a spectral distribution well described by a thermal
component.  The bright source RX J1856.5-3754 is compatible with a
perfect blackbody \citep{burwitz03}, with hints for a second cold
component \citep{sartore12}, detected also in optical and ultraviolet
analyses \citep{pons02}. For the faint source RX J0420.0-5022, the
initial claims for a line \citep{haberl04} were not confirmed by
later, deeper observations \citep{kaplan11}.

For the other XINSs, small deviations from perfect Planckian spectra
appear ubiquitously
\citep{haberl03,haberl04,vankerkwijk04,zane05,haberl06}. RX~J1308.6+2127
shows the deepest spectral feature among the XINSs, while in
RX~J0806.4-4123 and RX~J1605.3+3249, hints of weak spectral features 
appear at $\sim 300-400$ eV. In XINSs, the width of the gaussian
lines, $\sigma$, is generally $\sim 70-170$~eV, while the equivalent
width, $E_w$, ranges between $\sim 30-150$~eV. Note that the 5 XINSs
with claims for lines have similar temperatures, $kT_{bb}\sim 80-100$
eV, while the remaining two objects (with no apparent spectral
features) are the coldest ones, with $kT_{bb}\sim 45-60$ eV. Given the strength of the
dipolar magnetic fields of the XINSs, the physical origin of these
features has been proposed to be cyclotron absorption by protons.

CCOs are X-ray bright objects observed in the center of young
supernova remnants (SNRs), a class that is possibly comprising 
a few diverse objects. The bright CCO 1E~1207.4-5209 exhibits a complex
spectrum, phenomenologically fitted by a purely thermal continuum,
plus several very broad absorption lines. The central energies $E_0$ of these
features have been interpreted as the fundamental and upper harmonics
of the cyclotron absorption by electrons.

Among the radio-pulsars, the only two cases with reported absorption
lines are PSR~J1740+1000 \citep{kargaltsev12}, and the peculiar,
highly magnetized rotating radio-transient PSR~J1819-1458
\citep{mclaughlin07,miller13}, at slightly higher energies than what
observed in XINS, and similarly interpreted as cyclotron absorption
lines.

Among the most strongly magnetized pulsars, a few claims are present
in the literature, all tentatively interpreted as proton cyclotron absorption
lines. They were observed either in the persistent emission
(\citealt{rea03}, but see also \citealt{rea05}), or during bursts and outbursts
\citep{ibrahim03,tiengo13}, when the hard non-thermal emission is
enhanced. In these cases, the features lie in the non-thermal part of the spectrum, thus they are not considered in this work. Only in one magnetar, XTE\,J1810-197 \citep{bernardini09}, the
claimed feature is in a thermally-dominated part of the spectrum, at an
energy of $E_0\sim 1.1$ keV. The best-fit continuum model in this case
is a complex 3BB+line, however it can also be modelled by a BB
distorted by magnetospheric resonant cyclotron scattering
\citep{rea08}.

\section{Synthetic spectra}\label{sec:synthetic}

\subsection{Inhomogeneous temperature maps}

The coupled internal evolution of magnetic field and temperature in
the NS crust leads to inhomogeneous internal and surface temperatures
under the presence of strong magnetic fields. Magnetic fields in NSs
have often been approximated as dipolar; for this configuration, the
equator, where the magnetic field is tangential, is colder than the
rest of the NS surface, and the resulting temperature distribution is
symmetric with respect to the equator. However, in reality the
magnetic topology is likely to be much more complicated; in
particular, toroidal and multipolar internal fields are likely to play an important role for
at least a fraction of NSs. Strong dipolar toroidal
fields break the symmetry with respect to the equator, and produce
temperature distributions which yield small blackbody radii in
spectral fits \citep{perna13}. The magnetothermal simulations by
\citet{pons09} and \citet{vigano13} have showed that the temperature
distribution is very sensitive to the details of the field geometry,
which in turn depends to a significant extent upon the poorly known
initial conditions at NS birth.  Furthermore, small portions of the
surface are believed to be kept hot by the particle
bombardment from returning magnetospheric currents (rotationally
powered for cold, middle-aged and old pulsars,
\citealt{cheng80,halpern93,zavlin04}, or magnetically powered in
twisted magnetosphere of magnetars,  \citealt{beloborodov13}).


\begin{figure}
\centering
\includegraphics[width=.45\textwidth,angle=270]{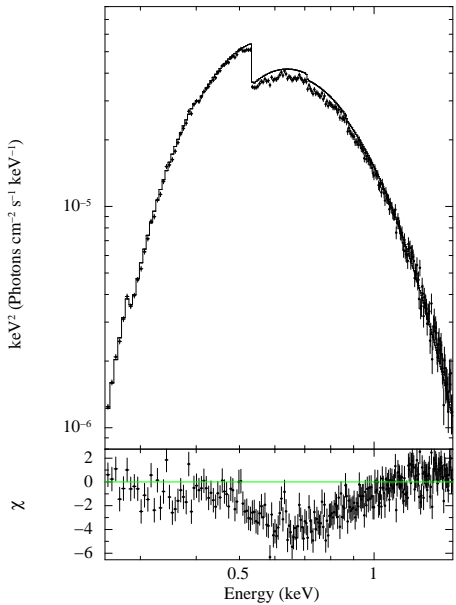}
\caption{Synthetic phase-averaged spectrum computed from the
  representative model B2, with the inclusion of an absorption column
  density $N_H=10^{21}$ cm$^{-2}$ (see
  Table~\ref{tab:synthetic_lines}). We show the unfolded spectrum,
  with residuals (in units of $\chi_\nu$), considering the model
  obtained by the best-fit BB+line model, and removing the line (in order
  to show its presence more clearly).}
 \label{fig:synthetic_nh}
\end{figure}

Given the theoretical uncertainties discussed above, it is not
possible to make specific predictions for what the temperature
distribution on the NS surfaces should be, and each case needs to be
studied on its own.  Therefore, for the general study of this work, we
have opted to explore a variety of prescribed surface temperature
profiles, without discussing their particular physical origin.  In
order to build them, we have used the most updated envelope models
(used in \citealt{vigano13} and adapted from \citealt{pons09}), with
prescribed values of crustal temperature and magnetic field.  In
Fig.~\ref{fig:temperatures} we show some of the models studied in
detail. The (un-redshifted) surface temperatures of all these models
are in the typical range $\sim 10-300$~eV. For simplicity, we divide
the models into three families. Family A, shown in the left panel,
shows anisotropies qualitatively expected from neutron stars with
crustal temperatures of $\sim 0.6-4 \times 10^8$~K, and surface
magnetic field $\sim 10^{13}$~G. Families B (middle) and C (right) add
a small hot spot to models from A; such localized heated regions could
have a variety of physical origins (i.e. particle bombardment, twisted
magnetic bundles), as mentioned above.

\subsection{Ray tracing and phase-dependent spectra}

The general relativistic, phase-dependent spectra are calculated using
the formalism developed by \cite{page95}. Due to light deflection by
the gravitational field of the star, a photon emitted at an angle
$\delta$ with the normal to the NS surface will reach the observer if
generated at an angle (with respect to the viewing axis)
\begin{eqnarray}
\theta_v(\delta) = \int_0^{\frac{R_s}{2R}}\, x\left[\left(1-{R_s\over 
R}\right)\left({R_s\over 2R}\right)^2-(1-2u)u^2 x^2\right]^{-\frac{1}{2}}\,\mbox{d}u,
\label{eq:teta}
\end{eqnarray}
\noindent
where $x\equiv\sin\delta$, $R_s\equiv 2GM/c^2$ is the Schwarzschild
radius of the star, and $M$ and $R$ are its mass and radius,
respectively. In this work, we employ $M=1.4M_\odot$ and $R=10$~km,
yielding a gravitational redshift  $(1-R_s/R)^{1/2}=0.766$.
The radius measured by a distant observer would be $R_\infty = R/({1-{{R_s}/{R}}})^{1/2}=13$ km. 

\begin{table*}
\begin{center}
\footnotesize
\caption{Results from our synthetic phase-average spectra for an orthogonal rotator ($\chi=\xi=90^\circ$), considering the energy band 0.1--10 keV. We report the simulated number of photons, the best-fit parameters, 1-$\sigma$
  errors, and $\chi_\nu^2$ values for the (absorbed, {\tt phabs})
  BB+line ({\tt bbodyrad*gabs}) model. We also show the comparison
  with the best-fit BB model, the related $F$-test, the value of $N_H$
  used in the synthetic spectra, $N_H^{\rm syn}$, and the value inferred
  from the fit, $N_H^{\rm fit}$. Radius and temperatures are reported as
  seen from an observer at infinity, at contrast with the physical
  unredshifted surface temperatures of
  Fig.~\ref{fig:temperatures}. Equally good fits are often obtained
  for a wide range of the Gaussian width, up to $\sigma\sim E_0$
  ($E_0$ being the line centroid): $\sigma$ has been fixed to the
  minimum value for which a good fit is found. For consistency, no
  error on the equivalent width ($E_w$) has been provided, since it is                                                                    
  very sensitive to the choice of $\sigma$. Note that, since we consider an orthogonal rotator, the shown values of PF represent the maximum value that each representative model can achieve for local isotropic emission.}
\label{tab:synthetic_lines}
 \begin{tabular}[ht!]{c c c c c c c c c c c c}

\hline 
\hline 
model & counts 	& $N_H^{\rm syn}$ & $N_H^{\rm fit}$ & $kT_{bb}$ & $R_{bb}$ & $E_0$ &	$|E_w|$ & $\sigma$ & $\chi_{\nu,bb+line}^2$ & $\chi_{\nu,bb}^2$ & PF \\
 	  & [$10^3$]	& $10^{20} {\rm cm}^{-2}$ & $10^{20} {\rm cm}^{-2}$ & [eV] & [km] & [eV] & [eV] & [eV] & & &  (\%) \\ 
\hline 
A1 	& 150	&  -	& -		& $44.5\pm 0.5$	&  $9.8\pm 0.3$	& $320\pm 20$	& 70	& 100	 & 1.05	& 1.53	& $3.2$	 \\ 
A2 	& 160	&  -	& -		& $53.5\pm 0.4$	&  $9.7\pm 0.2$	& $360\pm 30$	& 90	& 100	 & 0.99	& 1.40	& $3.1$	 \\ 
A3 	& 170	&  -	& -		& $61.6\pm 0.2$	&  $9.5\pm 0.1$	& $450\pm 30$	& 90	& 100	 & 0.98	& 1.45	& $3.0$	 \\ 
A4 	& 190	&  -	& -		& $77.4\pm 0.5$	&  $9.6\pm 0.1$	& $490\pm 20$	& 30	& 100	 & 1.04	& 1.57	& $2.8$	 \\ 
A5 	& 150	&  -	& -		& $91\pm 1$	&  $9.8\pm 0.1$	& $570\pm 40$	& 70	& 150	 & 1.00	& 1.32	& $2.8$	 \\ 
A6	& 180	&  -	& -		&$112\pm 1$	& $10.1\pm 0.2$	& $630\pm 50$	& 60	& 200	 & 1.05	& 1.41	& $2.7$	 \\ 
A7	& 120	&  -	& -		&$131\pm 1$	& $10.2\pm 0.2$	& $670\pm 50$	& 50	& 200	 & 0.95	& 1.31	& $2.6$	 \\ 
B1	& 30	&  -	& -		& $86\pm 1$	& $1.39\pm0.05$	& $440\pm 30$	& 50 	& 150	 & 1.10	& 1.42	& $90$	 \\ 
B2 	& 40	&  -	& -		& $106\pm 1$	&  $1.4\pm 0.1$	& $540\pm 30$	& 70	& 150	 & 1.07	& 1.37	& $94$	\\ 
B2 	& 210	& 1	& $0.5\pm0.1$	& $111 \pm 1$	&  $1.17\pm 0.04$& $670\pm 30$	& 50	& 150	 & 1.01	& 1.48	& $99$	 \\ 
B2 	& 250	& 10	& $9.0\pm0.3$	& $113\pm 1$	&  $1.09\pm 0.03$& $710\pm 30$	& 30	& 150	 & 1.11	& 1.31	& $100$	 \\ 
B3 	& 50	&  -	& -		& $123\pm 1$	&  $1.4\pm 0.1$	& $590\pm 30$	& 60	& 150	 & 1.03	& 1.38	& $96$	 \\ 
B4 	& 30	& 1	& $\lesssim 0.5$ & $91\pm 1$	&  $1.7\pm 0.1$	& $530\pm 30$	&140	& 150	 & 0.90	& 1.42	& $60$	 \\ 
B4 	& 30	& 1	& 1 (fixed)	& $83\pm 1$	&  $2.7\pm 0.1$	& $480\pm 20$	& 240	& 150	 & 1.04	& 1.42	& $60$	 \\ 
B4 	& 70	& 5	& $3.0\pm0.2$	& $98\pm 1$	&  $1.2\pm 0.1$	& $620\pm 30$	& 90	& 150	 & 1.04	& 1.52	& $83$	 \\ 
B5 	& 140	& 1	& $\lesssim 0.5$& $132\pm 1$	&  $1.3\pm 0.1$	& $690\pm 40$	& 30	& 150	 & 1.07	& 1.30	& $83$	 \\ 
B5	& 90	& 2	& $0.8\pm0.2$	& $128\pm 2$	&  $1.4\pm 0.1$	& $610\pm 40$	& 50	& 150	 & 1.12	& 1.36	& $89$	 \\ 
B5 	& 140	& 5	& $3.6\pm0.2$	& $128.6\pm 0.6$&  $1.4\pm 0.1$	& $620\pm 20$	& 60	& 150	 & 0.99	& 1.30	& $95$	 \\ 
C1	& 50	& 5	& $3.8\pm0.4$	& $126\pm$ 2	& $5.5\pm$0.4	& $810\pm$20	& 400	& 250	 & 1.34	& 2.68	& $29$	\\ 
C1	& 30	& 10	& $7.6\pm0.8$	& $133\pm$ 3	& $4.5\pm$0.5	& $890\pm$30	& 380	& 250	 & 1.12	& 1.87	& $35$	\\ 
C1	& 30	& 20	& $14.5\pm1.5$ 	& $140\pm$ 3	& $3.3\pm$0.3	& $900\pm$20	& 250	& 200	 & 1.13	& 1.50	& $44$	\\ 
C2	& 90	& 5	& $3.8\pm0.3$	& $134\pm$ 1	& $7.0\pm$0.1	& $880\pm$30	& 160	& 200	 & 1.18	& 2.01	& $16$	\\ 
C2	& 80	& 20	& $16\pm 1$	& $143\pm$ 3	& $5.5\pm$0.4	& $1020\pm 40$ & 180	& 250	 & 1.23	& 1.52	& $24$	\\ 
C3	& 50	& 100	& $78\pm1$	& $185\pm$ 2	& $1.8\pm$ 0.1	& $1560\pm$40 & 100	& 250	 & 0.97	& 1.31	& $58$	 \\ 
C4	& 180	& 5	& $4.3\pm0.2$	& $145\pm 1$	& $8.4\pm0.2$	& $870 \pm 30$ & 80	& 200	 & 1.03	& 1.55	& $11$	\\ 
C4	& 1000	& 50	& $47\pm1$	& $152\pm 2$	& $7.1\pm 0.5$	& $1030\pm 20$ & 80	& 300	 & 1.23	& 1.53	& $25$	\\ 
C4	& 640	& 70	& $59\pm1$	& $168\pm 1$	& $4.4\pm 0.1$	& $1440\pm 40$ & 50	& 200	 & 1.13	& 1.57	& $29$	\\ 
C4	& 330	& 100	& $87\pm1$	& $172\pm 1$	& $3.9\pm 0.1$	& $1500\pm 30$ & 60	& 200	 & 1.16	& 1.48	& $34$	\\ 

\hline
\hline
 \end{tabular}
\end{center}
\end{table*} 

As the star rotates with angular velocity $\Omega(t)$, the observer
measures the modulated photon flux
\begin{eqnarray}
F(E_\infty,\alpha)=\frac{E_\infty^2}{c^2h^3}\frac{R_\infty^2}{D^2}\;
\int_0^1 2x \int_0^{2\pi}
I[\theta_v(x),\phi_v,E]\,\mbox{d}\phi_v\mbox{d}x\,,
\label{eq:flux}
\end{eqnarray}
where $D$ is the distance, $E_\infty=E({1-{{R_s}/{R}}})^{1/2}$ is the
energy measured by a distant observer, the spectral function
$I(\theta_v, \phi_v, E)$ describes the dimensionless distribution of
the locally emitted photons, and $(\theta_v,\phi_v)$ are the
coordinates on the surface relative to the line of sight. Here we
assume the local emission to be isotropic and corresponding to a
blackbody distribution with temperature $T(\theta_v,\phi_v)$,
i.e. $I=(e^{E/k_BT} -1)^{-1}$.  

The temperature distributions are assumed to be axisymmetric (see
Fig.\ref{fig:temperatures}), and the magnetic axis is taken to
coincide with the axis of symmetry of the temperature distribution. If we
indicate by $\psi$ and $\xi$ the angles that the rotation axis makes
respectively with the line of sight and the magnetic axis, then the
angle between the line of sight and the magnetic axis, $\alpha$, is
given by the simple geometric relation
\begin{eqnarray}
\label{eq:alpha}
\alpha(t) = \arccos\left[\cos\psi\cos\xi+\sin\psi\sin\xi\cos\gamma(t)\right]\,,
\end{eqnarray}
where $\gamma(t)=\int \Omega(t) dt$ is the phase angle swept by the
star during its rotation (see Fig.1 in \citealt{perna08} for a graphical
representation of the viewing geometry). From Eq.~(\ref{eq:flux}), the 
phase-averaged spectrum can be readily
computed as an integral over the phase angle $\gamma$.
For simplicity, we focus our analysis on spectra computed with
$\psi=\xi=90^\circ$ (orthogonal rotator), while briefly discussing
the effects of the viewing geometry on the presented results.

\begin{figure}[t!]
\centering
\includegraphics[width=.45\textwidth]{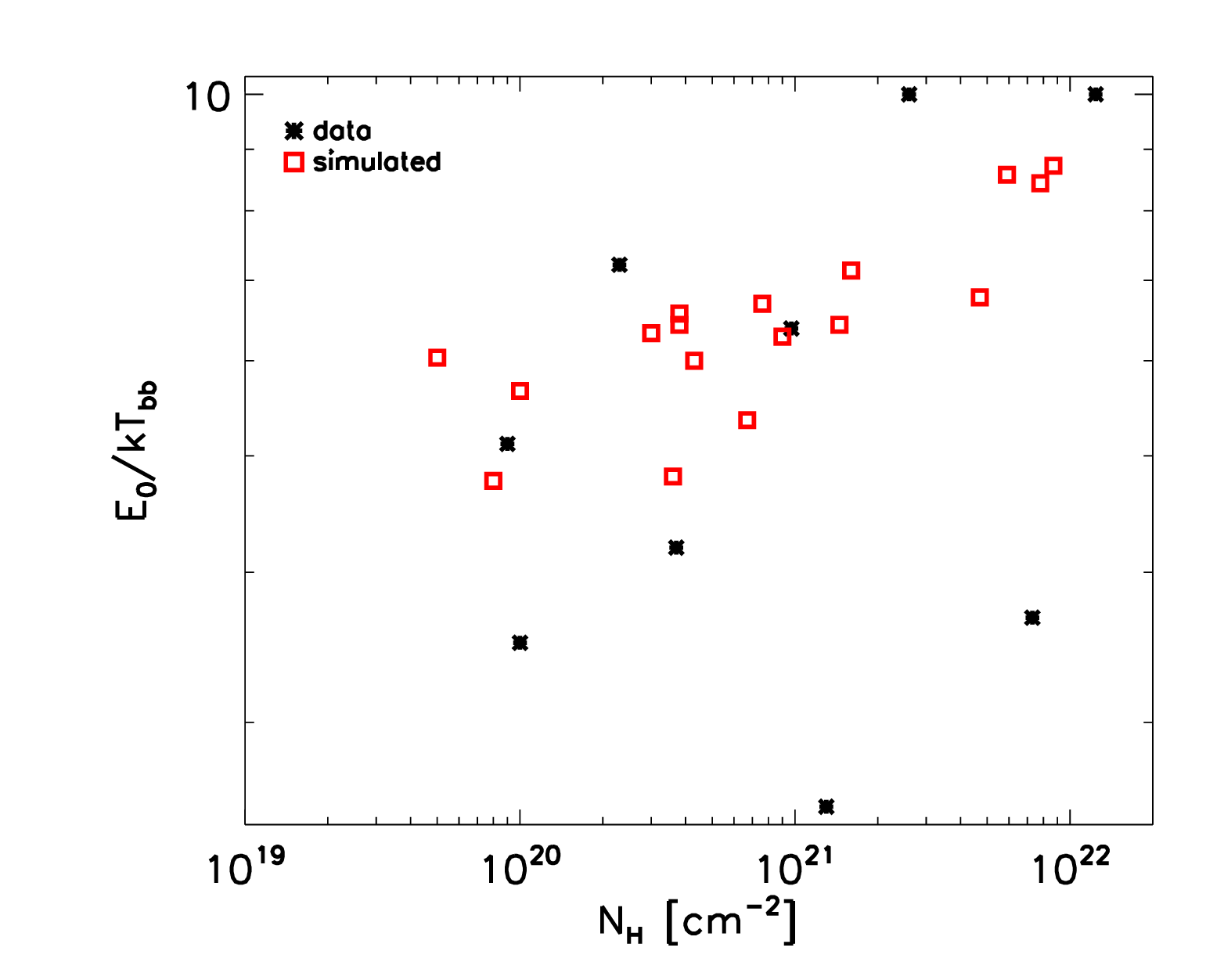}
\includegraphics[width=.45\textwidth]{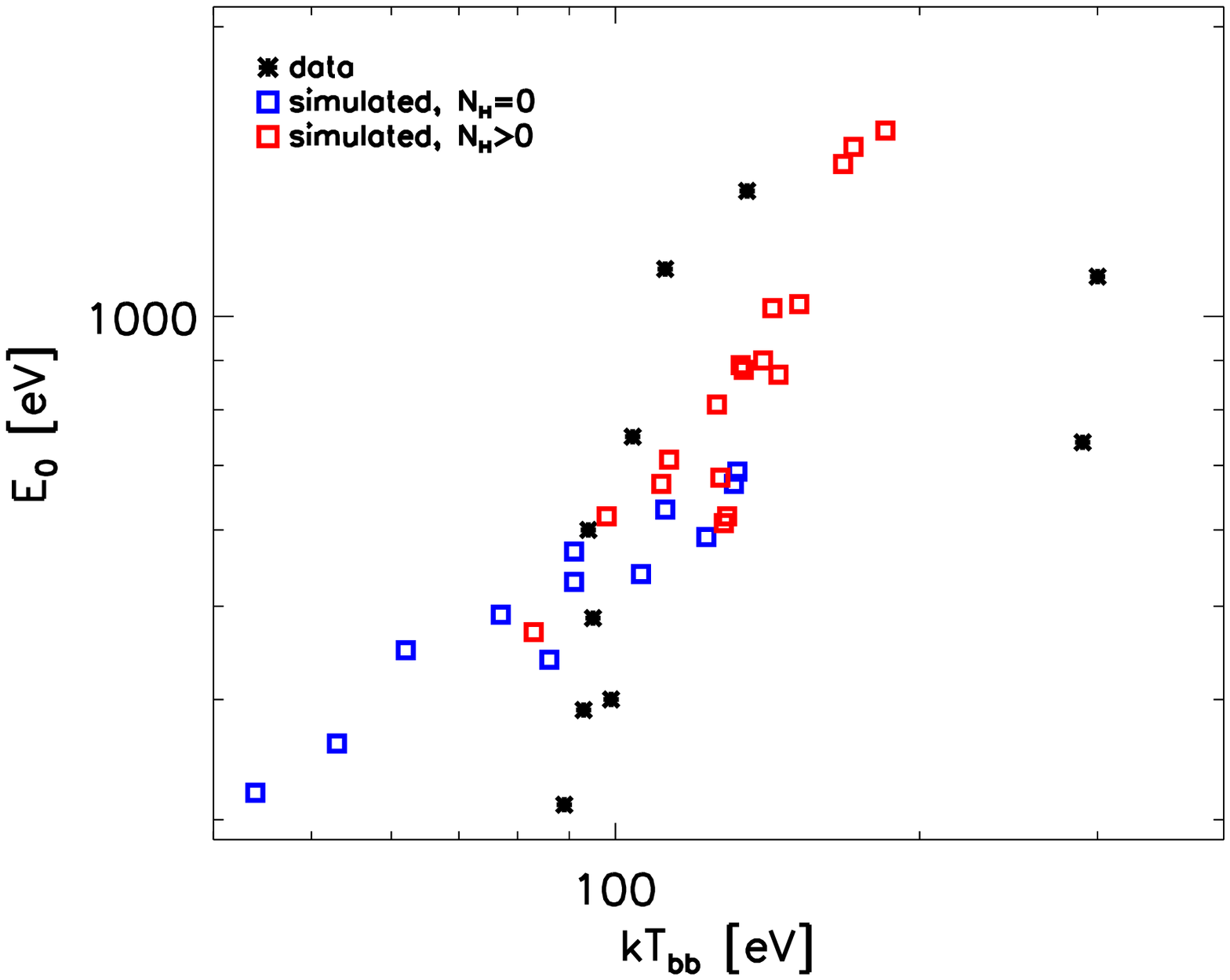}
\caption{Best fit parameters for BB+line model, applied to real data
  (black asterisks, see Table~\ref{tab:data_lines}), and synthetic
  spectra of Table~\ref{tab:synthetic_lines} (blue and red squares for
  models with $N_H=0$ and $N_H>0$, respectively). {\it Top:}
  correlation of $E_0/kT$ with $N_H$. {\it
    Bottom:} correlation between $E_0$ and $kT_{bb}$ (the hottest is
  considered in the case of real sources showing multi-T component).}
 \label{fig:correlations}
\end{figure}

\section{Simulated Spectra and Fit Results}\label{sec:results}

We imported the numerical, general relativistic spectra computed with
the various temperature distributions into {\tt Xspec}, as {\tt
  atable} models. We then simulated synthetic observed spectra, using
{\em XMM-Newton}/EPIC-pn response matrices in the {\tt fakeit}
procedure. For each model, we fixed the exposure time to be long
enough to obtain a number of photons that would allow to detect possible
spectral features.

Phenomenologically, the spectra of isolated NSs showing spectral lines
in their thermal emission are typically fitted by a model composed of a
thermal component (BB or atmosphere) plus a phenomenological
absorption model ({\tt gabs, gauss, cyclabs} in the fitting package
{\tt Xspec}, \citealt{arnaud96}). When a phase-resolved spectral
analysis is doable, spectral features can show phase-dependence (e.g.,
\citealt{deluca04,hambaryan11,kargaltsev12}).

For each simulated spectrum, we proceeded to the spectral analysis in
the standard way, considering energies between 0.1--10\,keV. We fitted
the simulated spectra with two models: (a) BB, (b) BB+line ({\tt
  bbodyrad*gabs} model in {\tt Xspec}), with the Gaussian width,
$\sigma$, fixed to a given value. Freezing $\sigma$ is a standard way
to proceed also in real observations (see, e.g., \citealt{haberl04})
when broad, weak features are present. As expected (see also
\citealt{perna13}), the spectra from nearly uniform temperature
distributions are always well fitted by a single BB.  For conciseness,
we do not discuss the trivial cases in detail. Hereafter we focus on
the most interesting configurations for our purpose, i.e. those
showing large temperature differences (at least a factor of $\sim$ 2)
between the coldest and the hottest regions (see
Fig.~\ref{fig:temperatures}).

We evaluated the systematic improvement in the fits obtained with the
BB+line model with respect to the pure BB, by looking at the shape of
the residuals and the values of $\chi_\nu^2$ ($\chi^2$ normalized by
the number of degrees of freedom) for the different realizations. We
confirmed the robustness of the results by finding, for each model,
the systematic appearance of a spurious line with similar parameters
in tens of different simulations.  By computing the contribution to
the flux of lines and continuum, we can also estimate the equivalent
width of the line, $E_w$. The latter, both in observations and in our
simulations, is typically tens or hundreds of eV. However, note that
equally good fits can often be found for similar values of $E_0$ but
different values of $\sigma$ (varying by factor of a few), or
different line models ({\tt gabs, gauss, cyclabs}). Since this
variability strongly affects the equivalent width, the values of the
latter should be taken as an order-of-magnitude estimate.  The results
of the fits to a variety of simulated spectra are summarized in
Table~\ref{tab:synthetic_lines}.

\subsection{Phase-averaged spectra}

We start by considering the representative case of the phase-averaged
spectra for model B2, simulated with $N_H=10^{21}$~cm$^{-2}$. A pure
BB fit gives an unacceptable $\chi_\nu^2$ and, more interestingly, it
shows systematic residuals. The addition of a broad absorption line
allows to better fit the data, at least by partly removing systematic
residuals. In Fig.~\ref{fig:synthetic_nh} we show the unfolded
spectrum with the residuals coming from the best-fit BB+line model
(removing the line for the sake of clarity), for a single realization
of the synthetic spectrum.

\begin{figure*}
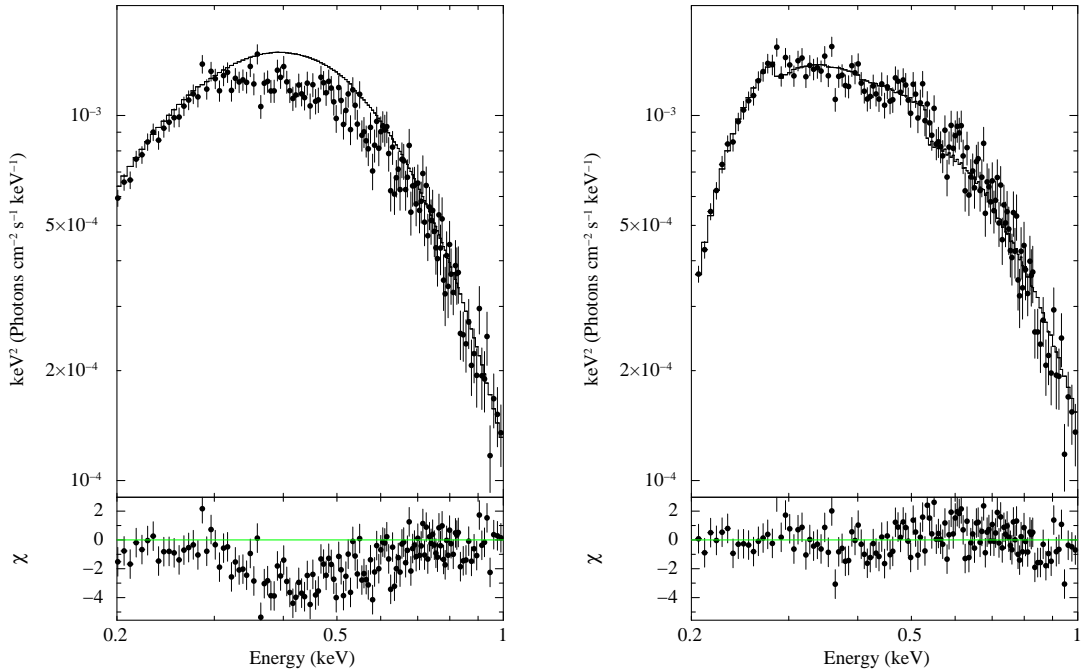

\centering
\includegraphics[width=.5\textwidth,angle=270]{images/rx0806_bb.ps}
\includegraphics[width=.5\textwidth,angle=270]{images/rx0806_model.ps}
\caption{Unfolded spectrum of RX~J0806.4-4123, and best-fit (with
  residuals in units of $\chi_\nu$) for a BB+line model
  ($\chi_\nu^2=1.36$, shown in the upper panel, with the line
  removed), and a model ($\chi_\nu^2=1.18$, bottom panel) generated
  with an inhomogeneous temperature distribution.   }
\label{fig:fit_0806}
\end{figure*}

The same qualitative behavior is seen in all the cases in which the
BB+line model fits better than the BB. The basic requirement needed to
create a feature in the spectrum is that the fluxes from the coldest
and hottest parts of the star be comparable. In these cases, the
spectral distribution, to first order, consists of two Planckian
shapes, whose temperatures and emitting areas represent an average
value of the coldest and hottest parts. The spectral feature appears
in the energy range between the two Planckian peaks. The larger the
difference in temperature, the deeper the feature.

In Table~\ref{tab:synthetic_lines} we have collected the fit results
from several model temperatures (among the ones explored) for which
the spectral feature appears. This sample is not meant to be a
complete and systematic exploration of the large space of possible
temperature distributions, but only an illustration of some of the
tested cases. For each of them, we show the best-fit parameters (with
statistical errors) of the BB+line spectral model for a single
realization, made with the indicated number of photons.  In general,
the central energy and the equivalent width are consistent with those
observed in the considered sample.

In both the BB and the BB+line models, the inferred temperature
$kT_{bb}$ corresponds to an average value of the redshifted surface
temperatures (see also \citealt{perna13}). The values of $kT_{bb}$ for
the BB and the BB+line models are compatible, while $R_{bb}$ can
increase by up to $\sim 10\%$ in the BB+line case, in order to account
for the flux absorbed by the line. An important point is that, in all
the cases, a 2BB fit (with absorption, if needed) fits well the simulated
data, as long as the counts are less than a few $\times 10^5$-$10^6$ (depending
on the model), in which case further small deviations can be resolved.

In the models with $N_H=0$, we find spurious lines for the cases in
which the majority of the surface is almost invisible though still
able to contribute to the low-energy spectrum, $E\gtrsim 0.1$\,keV. The
same effect is obtained with hotter models but with large absorption,
which again almost hides the cold component. In the case with
absorption included, $N_H$ is left free to vary, and usually the
best-fit absorbed BB (or BB+line) value underestimates the absorption
column by $\sim 20$-$30 \%$ with respect to the actual value used in
the simulation. As a matter of fact, the single BB model tends to
underestimate the flux coming from the colder part (which corresponds
to a larger, colder region), thus needing less absorption to fit the
data.

In Fig.~\ref{fig:correlations} we show the correlation of $E_0/kT_{bb}$
with $N_H$ (left), and of $E_0$ with $kT_{bb}$ (right), both in real data
(black) and in synthetic spectra (red). The spurious lines are centered
at $E_0 \sim 4$-$9~kT_{bb}$, corresponding to the tail of energy larger
than the Wien peak of the spectrum given by the best-fit BB model.  In
observations, some of the lines lie in the same regions, while others
are found at much larger energies, or much closer to the Wien peak. For
those cases, this implies that the origin of the
lines cannot be attributed to inhomogeneous surface temperatures.

Note also that selection effects limit the energy range in which lines can be
found: solid claims cannot be made for $E_0$ close to the lower
boundary of the energy domain. For large absorption, there are too few
photons in the region of low energy (few hundreds of eV), and therefore no
spectral details at low energy can be appreciated. The same applies to
the upper region of Fig.~\ref{fig:correlations} (left panel): thermal
photons in that part are too few for any deviation from a single
blackbody to be statistically significant.  Low values of $N_H\lesssim
10^{20}$ cm$^{-2}$ slightly affect only the very low energy part of
the spectrum, thus making $N_H$ poorly constrained and highly
correlated with $kT_{bb}$ and $R_{bb}$ (see, for instance, two
different fits for the model B4, with the same $N_H^{\rm syn}=10^{20}$
cm$^{-2}$ in Table~\ref{tab:synthetic_lines}).

In general, A-like models need a few $\times 10^5$ photons to make the
line detectable, while many of the B and C models need only $3$-$5\times
10^4$ photons: the stronger the anisotropy, the easier it is to
observe deviations from a pure BB, unless absorption is very large
(see some of the C models). Another difference lies in the inferred
radius from the fit ($R_{bb}$): while for the A models $R_{bb}$ is only
slightly smaller than the radius of the star (seen at infinity,
$R^\infty=13$ km), for the B and C models it is much smaller,
corresponding to the size of the hot spot (see
Fig.~\ref{fig:temperatures}).

Last, we computed, for each representative model, the predicted pulsed fraction
PF, defined as
\begin{equation}
PF=\frac{F_{\rm max}(\gamma)-F_{\rm min}(\gamma)}{F_{\rm max}(\gamma)+F_{\rm min}(\gamma)}\;,
\label{eq:pf}
\end{equation} 
where $F_{\rm max}$ and $F_{\rm min}$ are, respectively, the maximum and
minimum flux as a function of phase in a given energy band. We selected
the 0.1-2~keV band, above which the thermal flux becomes negligible for most models. All the
models are computed for an orthogonal rotator ($\psi=\xi=90^\circ$), which yields
the largest allowed value for the PFs for the considered emission model (blackbody).
The observed PF of each source puts some constraints on the allowed
range of angles, and this needs to be discussed case by case for specific objects.
We will do so in the following section for the source RX~J0806.4-4123.

\begin{figure}
\centering
\includegraphics[width=0.4\textwidth]{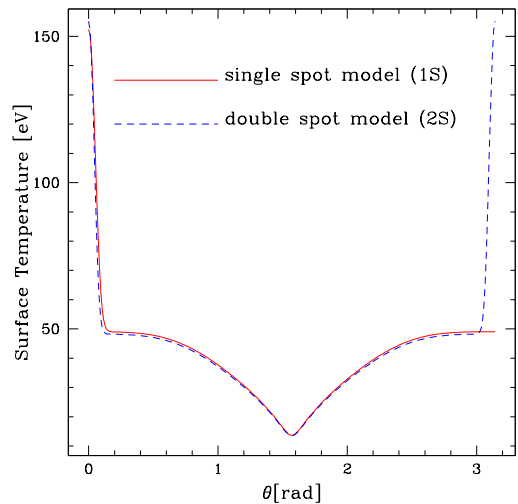}
\caption{Two possible temperature maps (single spot, with fit in Fig.~\ref{fig:fit_0806}, and double spot) that fit the phase-average spectrum of RX~J0806.4-4123.}
\label{fig:tprofiles}
\end{figure}

\subsubsection{An example: fit to J0806.4-4123}

In the previous subsection we have shown that the apparent lines in
the simulated spectra have $E_0$ and $E_w$ values similar to the
commonly observed ones. In the
following we directly test our models for a particular source,
J0806.4-4123. To this aim, we processed the 33.6 ks-long {\em
  XMM--Newton}/EPIC-pn observation of 2003 April 24$^{\rm th}$, using
{\tt SAS} version 12, and employing the most updated calibration
files.

\begin{figure}
\centering
\includegraphics[width=0.45\textwidth]{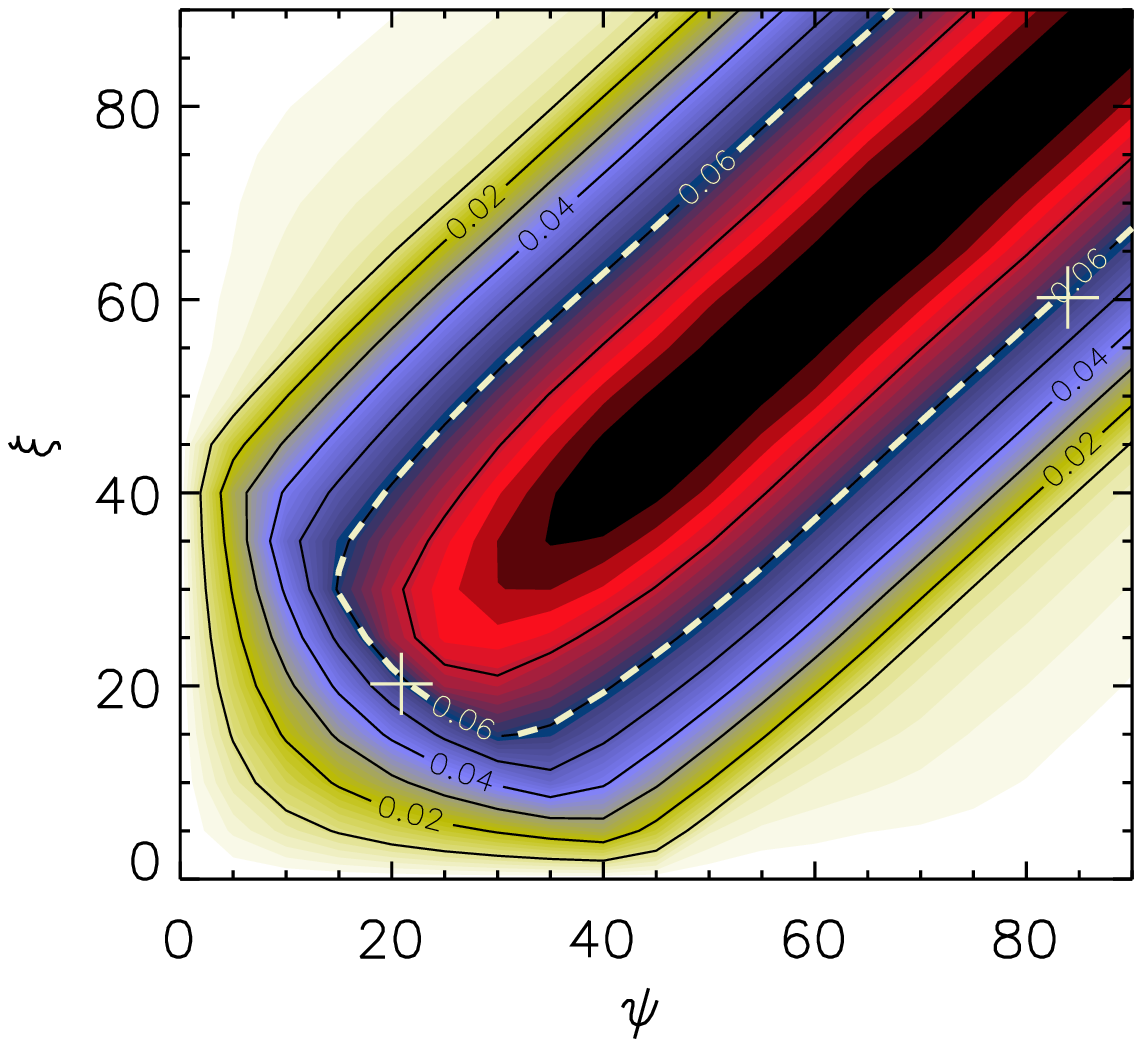}\\
\includegraphics[width=0.45\textwidth]{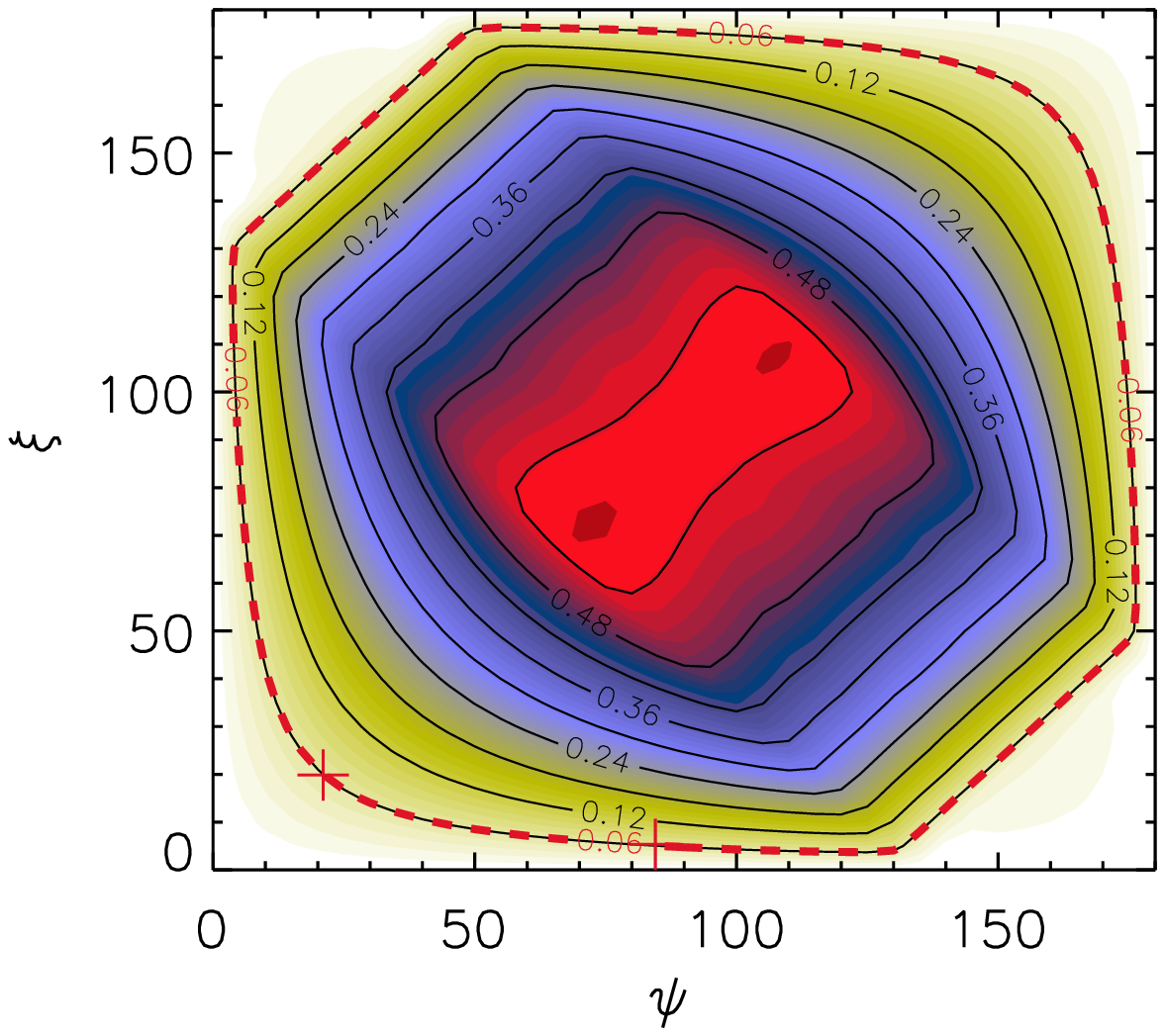} \\
\includegraphics[width=0.4\textwidth]{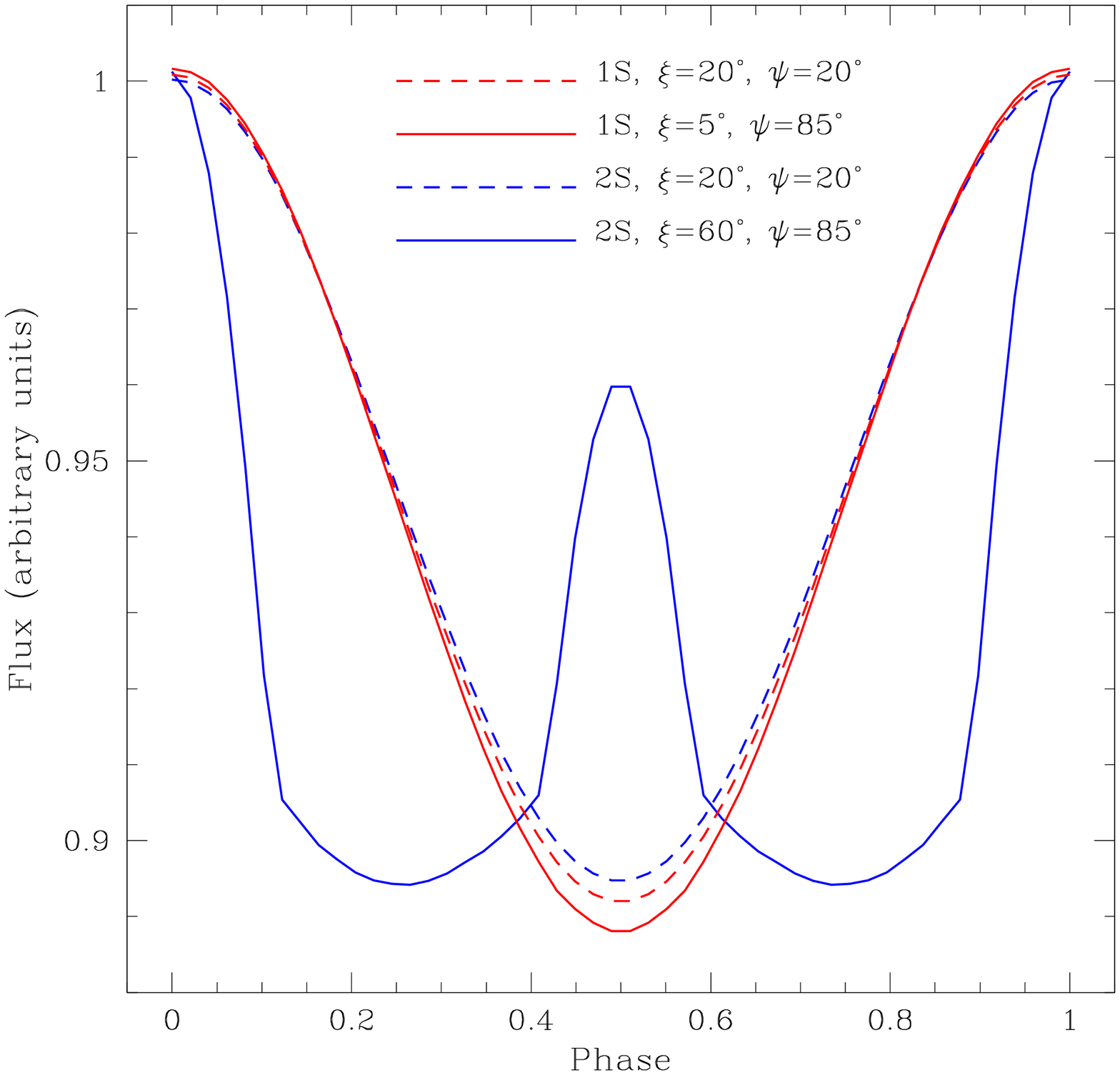}
\caption{PF map as a function of the angles $\psi$ and $\xi$ for
the models 2S (top) and 1S (middle) of Fig.~\ref{fig:tprofiles}. Given the symmetry of the
2S model, a restricted parameter space is shown. For each model, the bottom panel shows two pulsed profiles, each computed for a choice of angles which yield pulsed fractions consistent with the observed value of $\sim 6\%$.}
\label{fig:pulse_0806}
\end{figure}

In the left panel of Fig.~\ref{fig:fit_0806}, we show the unfolded
spectrum with the best-fit BB+line model ($\chi_\nu^2=1.36$) with $kT_{bb}=96$~eV,
$E_0=486$~eV, $E_w=30$~eV, and $N_H$ unconstrained. The line has been removed in
the plot for clarity. Several different temperature distributions (either with one or two hot spots)
are able to fit the spectrum better than the BB model and
the BB+line model. An example of one of such cases is shown in the right panel, with
$N_H=(4.2\pm 0.1)\times 10^{20}$ cm$^{-2}$, $\chi_\nu^2=1.18$\footnote{Note
that \cite{kaplan09b} further improve the fit by considering a model
with a BB and two absorption lines.}. Two of the typical temperature profiles that
fit the data (symmetric/asymmetric distributions) are shown in Fig. ~\ref{fig:tprofiles}.

\begin{figure*}
\centering
\includegraphics[width=.45\textwidth]{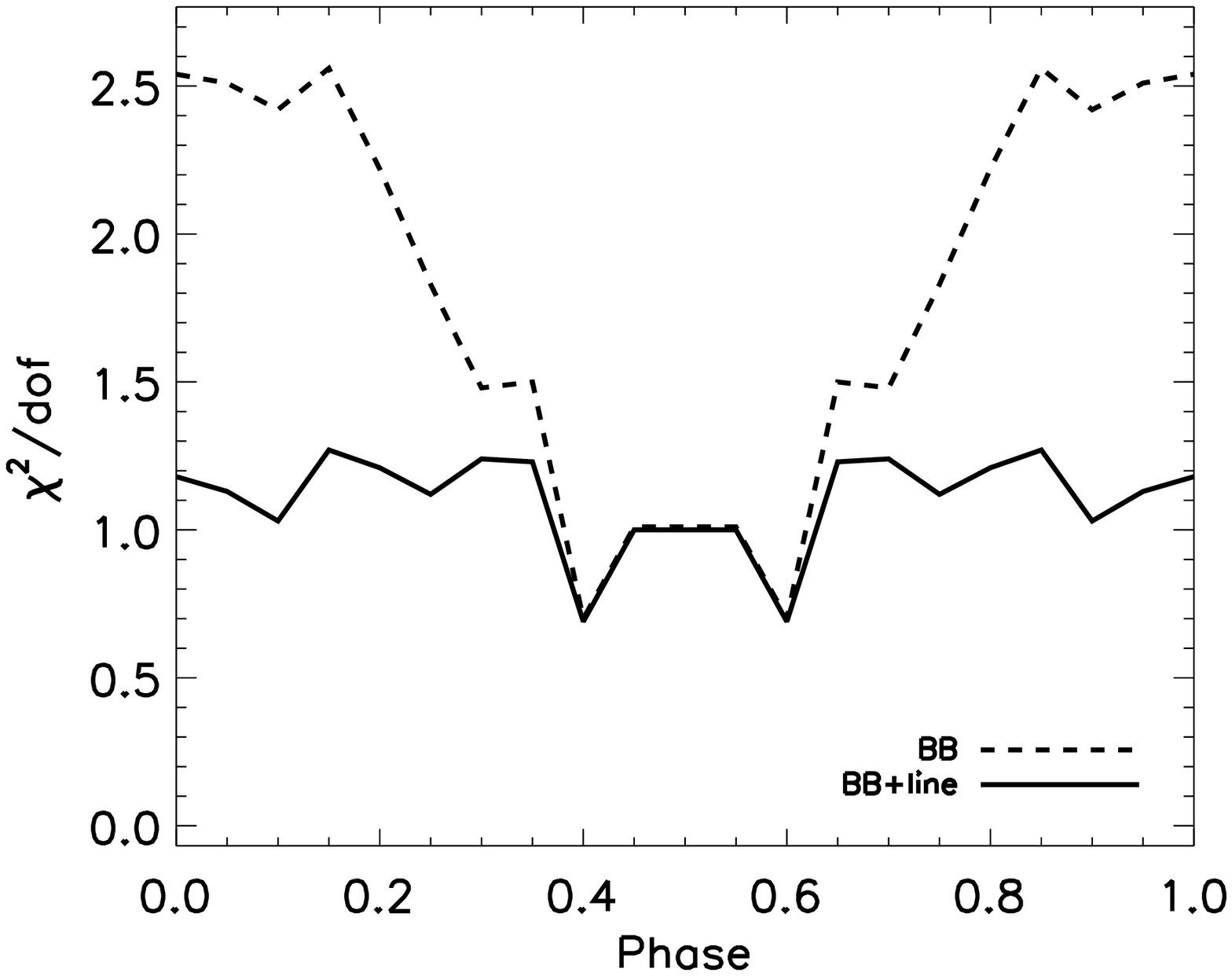}
\includegraphics[width=.45\textwidth]{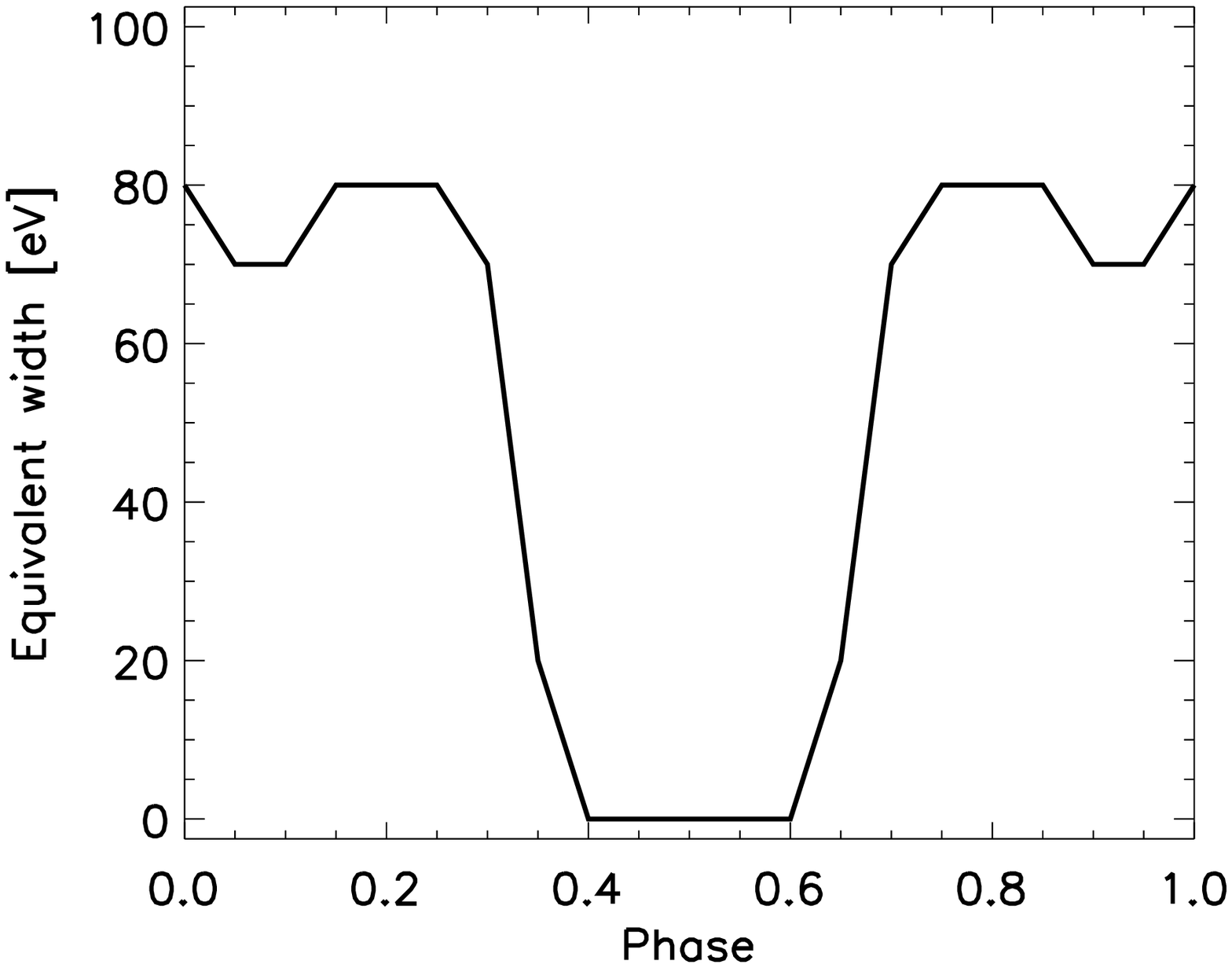}
\caption{Phase resolved spectra, for the model B2, without
  absorption. {\em Left}: $\chi_\nu^2$ of the best fitting BB (dashed)
  or BB+line model (solid) as a function of phase. {\em Right}:
  equivalent width, $E_w$, as a function of phase.}
\label{fig:phase_resolved}
\end{figure*}

An independent, important constraint on the underlying
  temperature profile is provided by the pulsed profiles. For the
  XINSs, they are generally single-peaked, with relatively low pulsed
  fractions. In the particular case of the source J0806.4-4123, the PF
  in the EPIC-pn is $6\%$ in the 0.12-1.2 keV band \citep{haberl02}. We hence explored the
  parameter space in the angles $\psi, \xi$ which can yield this
  modulation level, for both models. We note that, without the
  effect of interstellar absorption, the modulation level of the
  models in a wide energy band is low, even for an orthogonal
  rotator ($\psi=\xi=90^\circ$). This is because, as already
  discussed, the models that yield spectral features are characterized
  by comparable fluxes in the hot and colder regions (dominant at
  opposite phases).  However, for sources with high $N_H$, the colder component
  is largely absorbed, thus lowering the flux at the minimum of the pulsed profile and resulting in
  an increased PF in a wide band (see \citealt{perna00} for an extensive discussion of this topic).
This is particularly so for the single spot model.

Fig.~\ref{fig:pulse_0806} shows contour plots of the PF as a function
of the angles $\psi,\xi$ for the two models used for the fit of
J0806.4-4123 (Fig.~\ref{fig:tprofiles}). As expected, in each case
there is a range of viewing angles which yield values of the PF
consistent with the observed value of 6\% (thick dashed line in the
figure). In the single spot case, the profile is single peaked for all
the choices along this degenerate region. On the other hand, for the
double spot, some viewing geometries, even if consistent with the
observed PF, can however be ruled out as they would yield a double
peaked profile, unlike that observed. Among the range of angles
yielding PF$\sim 6$\%, we have selected, for the double spot, one
particular choice of angles yielding a double peaked profile, and
another yielding a single peaked one (favoured by the observations).
The single peak profiles require solutions to be close to the main diagonal,
with allowed angles in the range $\psi \approx \xi \approx 10^\circ-30^\circ$.  On
the other hand, solutions such as the one displayed with $\xi \approx 85^\circ$
and $\psi \approx 60^\circ$ would show a second (lower) maximum in
the pulsed profile.

Finally, we should note that the derived values of the viewing angles
should be taken with caution, for two reasons: first, here we have assumed
blackbody, perfectly isotropic emission, and realistic atmospheres are
known not to be such (e.g. \citealt{zavlin96}); second, a change of viewing angles implies, in turn, slighlty different phase-average spectra, and a non-trivial, iterative fine-tuning of both temperature distribution and viewing angles would be needed to achieve a fully consistent fit of all observed properties.

\subsection{Phase-resolved analysis}

The phase-resolved spectra strongly depend on the inclination and
viewing angles. In axial symmetry, the most extreme phase-variability
(and pulsed fraction, PF) is obtained when $\xi=\psi=90^\circ$ (orthogonal
rotator). The smaller the angles, the less distinguishable in phase
are the hot and cold regions, because all the hot spots/rings
are seen in most phases.

We analyzed the fake phase-resolved spectra for model
B2 with $\xi=\psi=90^\circ$, considering 20 different phase angles $\gamma$.
Again, we fit with a BB model, with or without a gaussian
absorption line with $\sigma$ fixed at 150 eV.

In Fig.~\ref{fig:phase_resolved}, we plot $\chi_\nu^2$ and the value
of $E_w$ as a function of density. The BB+line model fits
significantly better than the BB for all the phases at which the
northern hotspot is seen. On the other hand, around $\gamma=\pi$, the
hotspot is not seen, the temperature distribution is more homogeneous,
and a single blackbody fits well. The equivalent width remains
constant in phase, until the hot spot goes out of sight, and $E_w$
approaches zero. When the spurious line is detected, its centroid,
$E_0$, remains almost constant, within errors.  On the other hand,
models of family A show very little variability, because, being
axisymmetric, at least one of the two hot poles is always seen.

Therefore, asymmetric temperature distributions are characterized by a
strong phase variability in the equivalent width of the spurious line,
while the line centroid is barely affected. While these
phase-dependent properties may not be unique to spectral features due
to temperature inhomogeneities, they should be kept in mind in the
process of interpreting the origin of spectral features, case by case.

\section{Discussion}\label{sec:discussion}

We have computed synthetic X-ray spectra of NSs for several
inhomogeneous surface temperature maps and found that, in some cases,
the spectra present features compatible with broad absorption lines at
a high significance level. The distinguishing characteristic of the
temperature distributions leading to these spectral distortions is one
or more hot, small regions with a temperature larger by at least a
factor of about 1.5-2 than the average temperature of the rest of the
surface. The hot spot(s) and the cold parts should give similar
contribution to the flux; the poor sensitivity of the instruments at
low energy and/or the interstellar absorption, partially hide the
Planckian emission produced by the coldest regions. As a consequence,
the spurious line appears between the two Planckian peaks. In
  light of our analysis, we propose that the need for absorption lines
  in the fits to some isolated neutron stars may be a hint for
  inhomogeneous temperature distributions, but the opposite is not
  generally true: an inhomogeneous temperature does not always imply
  the appearance of a spectral line.  The NS in the SNR Kes79
  \citep{halpern10} shows strong hints for anisotropy (two BBs in the
  spectral fits, and/or a large pulsed fraction), but it does not
  require a line in the spectrum.

In our synthetic models, the values of the line model parameters
($E_0$ and $E_w$) are in the same ranges as the observed ones. The
spurious lines always appear beyond the Wien peak of the BB+line
model, and $E_0 \sim 4-9 ~kT_{bb}$. Most objects (e.g., XINSs and PSRs),
show a similar position of $E_0$. However, conclusions cannot be drawn
only on the basis of this property; rather, each object should be
considered separately. As an example, we have performed a fit to
RX~J0806.4-4123, finding several different temperature distributions
able to give acceptable fits, better than the BB+line fit. This shows
that, at least for this object, a temperature inhomogeneity can produce
deviations from a pure BB model similar to the observed ones. The gross features of  On the
other hand, other sources like CCO~1E1207.4-5209 and the XINSs
RX~J1308.6+2127 are cases for which the deviations from a BB spectrum
do not appear to be reproducible by inhomogeneous temperature
distributions. The exceptional spectral complexity of these sources
cannot be mimicked, even qualitatively, by any of the many surface
temperatures that we have tested.

In general, we remark that, given the wide space of parameters
of the temperature modeling, reconstructing the surface temperature
map is a highly degenerate problem. Moreover, for the purpose of
isolating spectral effects due to temperature anisotropy alone, here
we have used local blackbody emission rather than realistic,
magnetized atmospheric/condensed surface models. These are the reasons
for which we did not attempt individual fits to all the
sources. Rather, we have demonstrated that the contribution to
spectral features from inhomogeneous temperature distributions can be
significant, and hence this should be accounted for in combination
with more sophisticated emission models.

To conclude, we note that, while current observations still allow
several interpretations for the observed absorption features, the very
large effective area of the planned X-ray mission {\it Athena+}
\citep{nandra13} will provide top-quality spectra, with the
possibility of fine-tuning the spectral models and possibly discriminating
between different underlying physical mechanisms.

\section*{Acknowledgements}
This research was supported by NSF grant No. AST 1009396 and NASA grants
AR1-12003X, DD1-12053X, GO2-13068X, GO2-13076X
(RP); grants AYA 2010-21097-C03-02 (JAP); iLINK 2011-0303 (NR);
AYA2012-39303 and SGR2009-811 (DV, NR). NR is supported by a Ramon y Cajal fellowship and an NWO Vidi award.
DV thanks JILA (Boulder, CO, USA) for its kind hospitality during the
time that some of this work was carried out. We thank S.~Mereghetti, R.~Turolla and W.~Ho for their valuable comments, and the anonymous referee for the useful discussion.

\bibliography{lines}

\end{document}